\documentclass{elsart}
\usepackage{epsfig}
\usepackage{amssymb}

\begin{document}

\newcommand{\avg}[1]{\langle{#1}\rangle}
\newcommand{\be}{\begin{equation}}
\newcommand{\ee}{\end{equation}}
\newcommand{\beas}{\begin{eqnarray*}}
\newcommand{\eeas}{\end{eqnarray*}}
\newcommand{\bea}{\begin{eqnarray}}
\newcommand{\eea}{\end{eqnarray}}
\newcommand{\req}[1]{(\ref{#1})}
\def\bc{\begin{center}}
\def\ec{\end{center}}

\begin{frontmatter}

\title{Ecology of active and passive players and their impact on information selection}
\author{G. Bianconi$^1$, P. Laureti$^1$ , Y.-K. Yu$^{2,3}$  and Y.-C. Zhang$^1$}
\address{$^1$Institut de Physique Th\'eorique,Universit\'e de Fribourg P\'erolles, CH-1700 Fribourg, Switzerland \\
$^2$ National Center for Biotechnology Information, National Library of Medicine,
 NIH, Bethesda, MD 20894, USA\\
$^3$Department of Physics, Florida Atlantic University, Boca Raton, FL 33431, USA} 
\begin{abstract}
Is visitors' attendance a fair indicator of a web site's quality?
Internet sub-domains are usually characterized by power law
distributions of visits, thus suggesting a richer-get-richer
process. If this is the case, the number of visits is not a relevant
measure of quality. If, on the other hand, there are active
players, i.e. visitors who can tell
the value of the information available, better sites start
getting richer after a crossover time.
\end{abstract}                                                                                
\begin{keyword}
Internet \sep Information \sep Ranking
\PACS 89.65 -s\sep 89.75 -k \sep 89.20.Hh
\end{keyword}
\end{frontmatter}

Over the last decade, the increasing use of Internet and number 
 of websites have made available to a vast audience huge quantities
of information. Under such tremendous information overflow, two of the most  
 important problems are: ({\it i}) what is the characteristic of such systems?  
 and ({\it ii}) how does one find relevant information in the ocean of the WWW? 
The challenge of finding relevant information is not new, well before the advent of Internet.
 Take the bestsellers for books, some readers would buy books which are high on the rank list, thus
 enhancing the highly ranked books standing; some other would buy only if it genuinely passes their 
 criteria, regardless the ranking. The former group of readers is said to be passive and the 
 latter active. What is true for books must also be valid for movies and consumer products and services, 
 political candidates, and myriad of things in our modern social economic life.
 
 It's clear that if all of the players are passive, what happens would be a richer-get-richer
 scheme, and the Barabasi-Albert model provides the paradigmatic example
 for a broad class of phenomena \cite{LINKED}. Our work below is an attempt to introduce a 
 certain number of so-called active players who know better what they want, not duped by 
 the possibly misleading signal, find the targeted object and her action hence can enhance
 the ranking of the otherwise ignored item. We are particularly interested in the ratio of passive to active
 players and the outcome of information selection capability. In the real world most of us are passive and on a few occasions
 we are active. This is because that in the normal conduct of daily life, we cannot be expert in all subjects,
 we necessarily rely on others' information selection capability to find what we want. So in this connected
 world we want to model that on each specific niche there are some active players, while all others are passive, following
 a mechanism akin to the BA preferential attachment.

Like many other
quantities characterizing social networks, web site attendance \cite{CuBeCr95}  and 
connectivity \cite{BAH} seem to be power-law distributed. The BA model \cite{BA} 
is believed to explain the fundamental mechanism underlying evolving networks, but
does not account for the selection of valuable information. This can be achieved
by assuming web sites have a {\em scalar} intrinsic quality which people can, 
to a certain extent, take into account besides their popularity \cite{BiBa01}.
 Nevertheless, we are aware that 
 approximating the quality of a web page by a scalar is only adequate
 when comparing web pages under the same category. Otherwise,
  it will be as inappropriate as providing an absolute ranking  
 among, say, physicists and apples. 
 The motivation of the current work is based upon two fundamental 
 observations that have not received
 much attention so far. 

The first one is: 
{\em the correlation between the popularity of a web site and its
 quality emerges from the interplay of heterogeneous visitors}.
In fact it is known that old sites do enjoy an edge over new and less popular 
ones. 
That is due to the fact that most visitors are {\it passive}, i.e. they are easily
influenced by advertising, word of mouth or a web site's rank in search engines.
There are, on the other hand, some people for whom a given information
is, for some reason (it may concern their core business or their main hobby), 
of great importance. 
They will spend a great deal of resources (their money, time and capability) 
searching for it,
and they will reward good sites regardless of how famous they are. These
 {\it active} visitors,
although a minority, are responsible for the selection, and eventual popularity,
 of good web sites. The ecology of active and passive players has already been dealt with
in different contexts \cite{Andrea}.
The second observation is: 
{\em there is no intrinsic reason why social networks should display a power 
law with a given exponent forever}, 
we have no control over the changes 
that the parameters governing it may undergo. Furthermore, models aiming
 to describe 
a network do not need be asymptotically scale free, but they might have a crossover 
between different regimes.

 Because we would like to address the {\it quality} issue and are 
 aware of the limitation of scalar representation for quality, 
  we will study a simple stochastic model that is applicable to 
 attendance statistics of web pages under the same category, i.e.,
 the quality of each web page will be represented by a scalar 
 quantity and the active visitor perceives the quality indicator 
 of each site. To model a larger system with various categories,
 we have studied a simplified situation where only passive
 visitors are considered. This is to bypass the more complicated mathematics
 needed to model the cross-category quality assessment but still hope to 
 capture the early-time statistical properties of the network.  
 The omission of active visitors is not a severe drawback here because  
 each visitor can probably be active in only a few categories and
 the statistical properties before the active opinions 
 become amplified are mostly influenced by  queries from passive
 visitors. 

The paper is therefore organized as follows. In section one 
we describe a simple stochastic toy model of web page attendance. Players
 can be either
active or passive, the precision of their activity being determined by an external
 temperature.
The model displays a power law distribution of web page visits and
a crossover between two regimes, where the choices of active players
are more or less influential.
In section two we shall analyze particular mean field instances of 
the model, 
with web sites' qualities respectively delta and uniformly distributed, where the
 stochastic part has 
been averaged out. In section three we sketch the analytical solution of the 
original model.
In section four,  while suppressing the presence of active visitors,  
 we model the system of multiple categories by using a 
 hierarchical geometry.  
 The final section documents our conclusion and some remarks. 

In the following we shall use interchangeably the terms web site and web page. 
They are
often referred to as nodes of a graph, in the language of networks.

\section{Active and passive players: a stochastic model}
We would like to describe now a simple model of web page attendance in a local
network. 
The number of web sites $N$ is fixed in the time-frame of the visits dynamics.
Each one of them is endowed with an intrinsic quality ${\epsilon_i}$, distributed
according to a given function $p(\epsilon)$. Such a scalar fitness is suitable to compare
sites belonging to the same domain.
At each time step $t$ a new player places a query. 
In order to account for visitors' heterogeneity in a minimal model, we consider the case in 
which players can be included in two different classes: active and passive.
With probability ${\rho}$
the player will be active, driven by the quality of web sites. 
Although his decision be affected by a noise, 
he visits site $i$ with average probability
\be\label{pi}
p_i=\frac{e^{\beta\epsilon_i}}{\sum_j e^{\beta\epsilon_j}},
\ee
where $\beta$ is an external parameter which plays the role of an inverse temperature.
Active players may be thought of as experts of the domain under consideration.
With complementary probability $(1-\rho)$, on the other hand, the player
is passive, driven by the popularity of web sites. His probability $g_i$ of
visiting site $i$ follows linear preferential attachment:
\be\label{gi}
g_i(t)=\frac{n_i(t)}{\sum_{j=1}^N n_j(t)},
\ee
where $n_i(t)$ is the number of visits to site $i$ at time $t$.
Notice that $\sum_{j=1}^N n_j=t$ for we have one visit per time step.

The following stochastic equation is intended to mimic the model
just described:
\be \label{eq}
n_i(t+1)- n_i(t) = (1-\rho)\frac{n_i(t)}{t} + \rho \xi_i(t) \\
\ee
with initial conditions
\be\label{noiseic}
n_i(0)=0, \; n_i(1)= \xi_i(0).
\ee
The stochastic noise $\xi$ is distributed as follows:
\be \label{noiseprob}
p(\xi_i)= p_i \delta(\xi_i -1) + (1-p_i) \delta(\xi_i).
\ee
The first term on the rhs of \req{eq} accounts for the ``richer get richer''
phenomenon due to passive players using preferential attachment \req{gi}.
The second term is stochastic and describes the behavior of active players.
On average they employ probability \req{pi}, but they are only allowed to
pick one site when they come into play. Therefore the noise term must be
normalized as follows:
\be \label{noisesum}
\sum_{i=1}^N \xi_i(t) =1, \; \forall t.
\ee

\section{Mean field results}
If we average equation \req{eq} over the noise, making use of \req{noiseprob},
and take the continuous time limit,
we obtain the corresponding mean field evolution equation. We shall analyze it 
for two significant
instances of the distribution of web sites' quality $p(\epsilon)$. 

\subsection{Random choice}
Let us consider a network where all web sites share the same quality value $\epsilon_i=\epsilon$.
Since, in this case, the preference probability of active players \req{pi}
becomes $p_i=1/N$, they actually place random queries. The evolution equation
reads in this case:
\be\label{uno}
\frac{\partial n_i}{\partial t}=(1-\rho)\frac{n_i}{t}+\frac{\rho}{N},
\ee
with initial conditions $n(t,t_i)=0$ $\forall t<t_i$ and
$n_i(t_i,t_i)=1$. Here $t_i$ is the time of the first visit to site $i$.
For $t>t_i$ the solution is:
\be\label{solution}
n_i(t,t_i)=\frac{t}{N}+\left(1-\frac{t_i}{N}\right)\left(\frac{t}{t_i}\right)^{1-\rho}.
\ee
The statistics of the first-time $t_i$ is well described by the probability that a site $i$ is not 
searched by active players up until time $t_i$, i.e.
$$
P(t_i)=\frac{\rho}{N}\left({1-\frac{\rho}{N}}\right)^{t_i}\simeq \frac{\rho}{N}e^{-\frac{\rho}{N}t_i},
$$
where the last formula holds in the large $N$ limit.
The average number of visits to a site is given by
\bea
<n(t,t_i)>_{t_i}&=&\int n(t,x)P(x)dx \nonumber\\
&=&\frac{t}{N}+\left({\frac{t\rho}{N}}\right)^{1-\rho}\left[-\frac{1}{\rho}\Gamma({\rho+1})+\Gamma({\rho})\right] \nonumber\\
&=&\frac{t}{N}\nonumber,
\eea
as expected.
Upon separating the terms due to the action of active and passive
players in \req{solution} and equating them, one finds
the crossover time
\be\label{crossover}
t_c \simeq (1+\rho)^{(1/\rho)} N/\rho.
\ee

We have performed numerical simulations of this model, gathering data
at  time $T=500$.
\begin{figure}
\centerline{\epsfxsize=3.0in \epsfbox{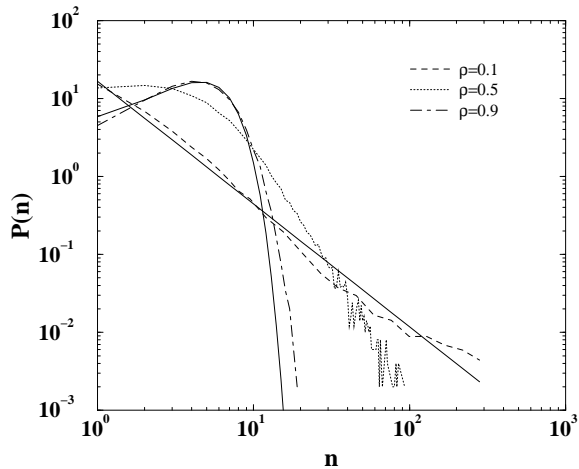}} 
\caption{Simulation results of the model described in \req{uno} for $\rho=0.1,0.5,0.9$ with $N=100$ web pages at time $T=500$. 
Data are averaged over $500$ runs. The solid lines are fits to the data: power law for $\rho=0.1$ and gaussian for $\rho=0.9$.} 
\label{cinquecento.fig}
\end{figure}
In fig. $\ref{cinquecento.fig}$ we plot the probability distribution of visits for  three different
values of $\rho=0.1,0.9,0.5$, whose corresponding crossover times 
\req{crossover}
are $t_c=26 N,2.5 N, 4.5 N$. After $T=500$ time steps,
therefore, the system of $N=100$ sites is expected to be
found, respectively, in the passive dominant, in the active dominant
and around the transition region for the three $\rho$ values chosen.
In fact we observe that for $\rho=0.1$, when passive players  are the majority, the density $P(n)$ 
of sites with a given number $n$ of visits decreases as a power-law of $n$, whereas  
for $\rho=0.9$, when active players are the majority, it follows a gaussian distribution.
Finally, for $\rho=0.5$, we observe a more complex intermediate behavior.

\subsection{Uniform quality distribution}
\begin{figure}
\centerline{\epsfxsize=4.5in \epsfbox{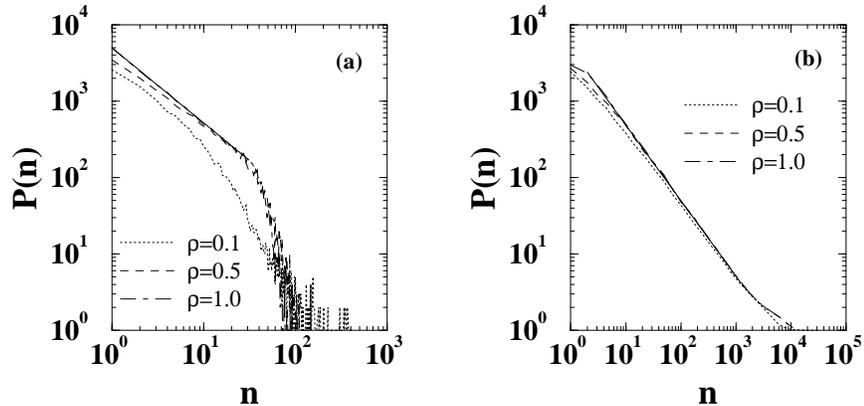}} 
\caption{Probability of having $n$ visits at times $T=500$ {\bf(a)} and  $T=10^5$ {\bf(b)},
following equation \req{due}.
Simulations are performed for a system of $N=100$ web pages, 
with $\beta=10$ and $\rho=0.1,0.5 1.0$.
In both graphs the power-law 
expected behavior for $\rho=1.0$, $P(n)\sim1/n$ \req{zipf} is represented by a solid line. } \label{pn.fig}
\end{figure}

We shall now analyze the continuous time limit of equation \req{eq}, averaged with respect to the noise distribution \req{noiseprob}, 
with $\epsilon_i$ uniformly distributed between zero and unity.
The corresponding evolution equation reads:
\be\label{due}
\frac{\partial n_i}{\partial t}=(1-\rho)\frac{n_i}{\sum_j n_j}+\rho \frac{e^{\beta\epsilon_i}}{\sum_j e^{\beta\epsilon_j}},
\ee
with initial conditions $n(t,t_i)=0$ $\forall t<t_i$ and
$n_i(t_i,t_i)=1$. For $t>t_i$ the solution reads
$$
n_i(t,t_i)=p_i t + (1-t_i p_i) \left(\frac{t}{t_i}\right)^{1-\rho}.
$$
As in the random case, this model displays two different regimes.
The crossover time can be similarly computed:
\be\label{tcross}
t_c \simeq N \; \frac{(1+\rho)^{(1/\rho)}}{ \rho}\;\frac{1-e^{-\beta}}{\beta}.
\ee
For $t \gg t_c$ the dynamics is
dominated by active players, who select valuable web sites.
In the $\rho=1.0$ limit, in particular,
we only have active players in the system.
The average number of visits to a given web page $i$ reads
$$
n_i(t)=\frac{e^{\beta \epsilon_i}}{\Lambda}t,
\label{n_i_epsilon}
$$
where $\Lambda=\sum_{i=1}^N e^{\beta \epsilon_i} $.
The probability distribution of the number of visits satisfies the relation
$
P(n)dn=p(\epsilon)d\epsilon
$
which, together with $(\ref{n_i_epsilon})$, gives
the power law distribution
\be\label{zipf}
P(n)\sim \frac{1}{n}.
\label{pl.eq}
\ee
We have simulated this dynamics for different values of $\beta$ 
and in the entire range of the $\rho$'s.
In Fig. $\ref{pn.fig}$ we report distribution $P(n)$ at 
small times $T=500$ (graph(a)) and at longer times $T=10^5$ (graph(b)). The solid lines plot the theoretical  
curve $(\ref{pl.eq})$ expected to be valid in the $\rho=1.0$ limit.
\begin{figure}
\centerline{\epsfxsize=4.5in \epsfbox{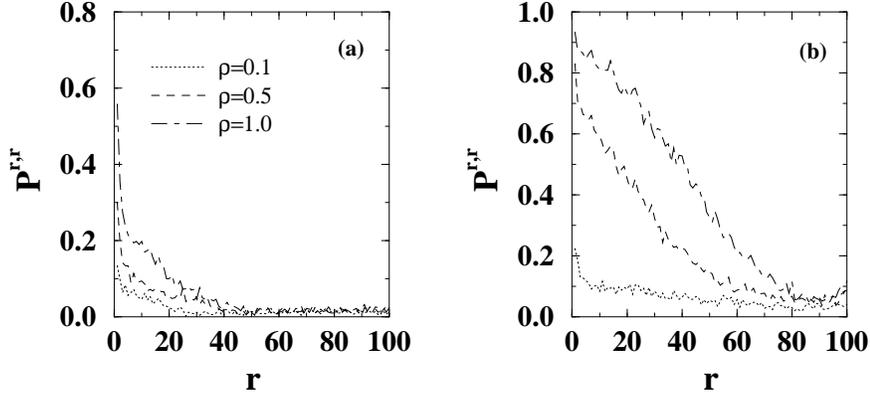}} 
\caption{Probability $P^{r,r}$ versus $r$ in model \req{due}.
Data are shown for a system of size $N=100$ web pages, $\beta=10$ and number 
of visits $T=500$ and $T=10^5$ respectively for graph {\bf(a)} and graph {\bf(b)}. Average taken 
over $500$ runs.} \label{Pr_r}
\end{figure}
\begin{figure}
\centerline{\epsfxsize=4.5in \epsfbox{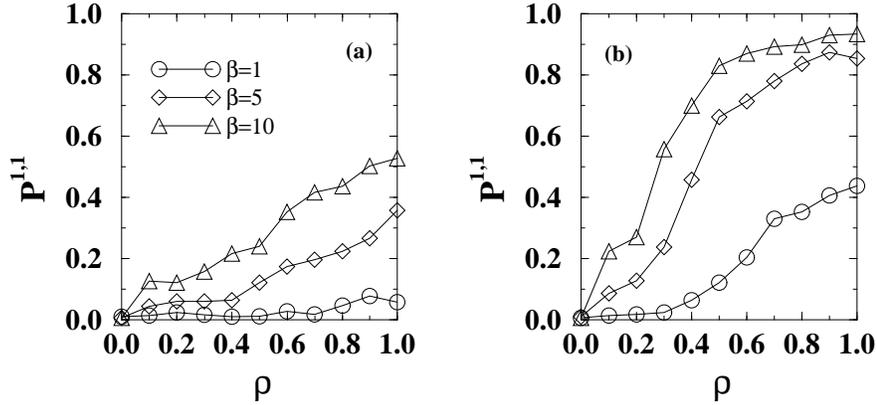}} 
\caption{Probability $P^{1,1}$ that the most visited web page is the fittest one in model \req{due}.
Simulations have been performed over a
system of $N=100$ web pages at $T=500$ (graph {\bf(a)}) and $T=10^5$ (graph {\bf(b)}) number of visits. 
Probabilities are calculated over $500$ runs. } \label{P1,1t500.eps}
\end{figure}

In order to measure how much the differences in the quality of web pages is
reflected in the  number of visits they receive, we have ranked them 
in order of decreasing number of visits and of increasing quality.
We have then measured the probability $P^{r,r}$ a web page ranked $r$ in  the ranking 
of the number of visits would coincide with the web page ranked $r$ in the quality ordering.
In order to measure $P^{r,r}$ we have ran the visit dynamics for $500$ time steps 
and we have calculated the number of times the web pages ranked $r$ in the two possible 
orderings were the same. 
In fig. $\ref{Pr_r}$ we plot simulation data
for different $\rho$'s at number of visits $T=500$ and $T=10^5$.
In fig. $\ref{P1,1t500.eps}$ we plot the probability that 
the web page that takes the maximal number of visits is the best one, i.e. $P^{1,1}$, as a function of $\rho$.

\section{Analytical results for the stochastic model}
Now we turn to our original model, described by equation $(\ref{eq})$ with
initial conditions \req{noiseic},
and try to solve it making use of methods similar to those outlined in references \cite{YK}.
To this end it is useful to define the
generating functions:
\bea \label{Gidef}
G_i(\lambda)=\sum_{t=1}^{\infty} \lambda^t g_i(t) \\ \label{Xidef}
\Xi_i(\lambda)=\sum_{t=1}^{\infty} \lambda^t \xi_i(t),
\eea
where $g_i(t)=\frac{n_i(t)}{t}$.
In fact, if we multiply $(\ref{eq})$ by $\lambda^t$ and sum over $t$,
we obtain
\beas
(1-\lambda) \partial_{\lambda} G_i(\lambda) = 
(1-\rho) G_i(\lambda) + \rho \Xi_i(\lambda) + \xi_i(0),
\eeas
which admits the following formal solution
\be \label{Gisol}
G_i(\lambda)=\xi_i(0)\frac{1-(1-\lambda)^{1-\rho}}{1-\rho} +
\rho (1-\lambda)^{1-\rho} \int_0^{\lambda} dx \frac{\Xi_i(x)}{(1-x)^{2-\rho}}.
\ee
Comparison between the small $\lambda$ development of \req{Gisol} 
and definition \req{Gidef}, yields
$$
g_i(t)=\sum_{s=0}^{t-1} \xi_i(s) A_{\rho,t}(s),
$$
with coefficients
\bea\label{azero}
A_{\rho,t}(0)&=&\frac{(-1)^{t+1}}{t!} 
\frac{\Gamma(\rho +t-1)}{\Gamma(\rho)}\\ \label{aesse}
A_{\rho,t}(s)&=&\rho 
\left[ 
\frac{\Gamma(s+1) \Gamma ( \rho +t-1)}
{\Gamma (t+1) \Gamma ( \rho +s)}  
\right].
\eea

\subsection{Probability distribution}
We are now able to write a formal expression for $P_t(\{ g_i \})$,
the probability density function of having a certain set of $g_i$-s ($i=1,2,...,N$)
at time $t$:
\be \label{Pdef}
P_t(\{ g_i \}) =  
\int \delta\left( g_i(t)-\sum_{s=0}^{t-1} \xi_i(s) A_{\rho,t}(s) \right)
\prod_{t'=0}^{t-1} 
\prod_{j=1}^{N} 
d\xi_j(t') p\left( \xi_j(t') \right) \,
\delta_{kr} \left( \sum_{l=1}^{N} \xi_l(t') -1 \right).
\ee
We can now employ the Fourier representation of the Dirac delta function
$$
\delta\left( g_i-\sum_{s=0}^{t-1} \xi_i(s) A_{\rho,t}(s) \right) =
\lim_{\alpha\to0} \frac{1}{2\pi} \int_{-\infty}^{\infty} dk_i 
\exp\left[ -\alpha k_i^2 + \imath k_i \left( g_i-\sum_{s=0}^{t-1} \xi_i(s)  A_{\rho,t}(s)\right) \right]
$$ 
inside \req{Pdef}. Thus we can separate the noise and integrate it out:
\beas
P_t(\{ g_i \}) =  \left[ \frac{1}{2\pi} \prod_{k=1}^N (1-p_k) \right]^t 
\prod_{l=1}^{N} \int dk_l \exp(-\alpha k_l^2 + i k_l g_l) \\
\prod_{s=0}^{t-1} \sum_{h=1}^{N} \frac{p_h}{1-p_h} \exp(-\imath k_h A_{\rho,t}(s)).
\eeas
By rewriting 
the last term of the integrand as
a product of sums (instead of a sum of products) and 
taking the limit $\alpha \to 0$, the former expression becomes:
\bea \nonumber
P_t(\{ g_i \}) =  \sum^{1,N}_{k_1, k_2, ..., k_N; \sum_{i=1}^{N} k_i=t}
\left( 
\prod_{i=1}^N 
\frac{e^{\beta \epsilon_i}}{\sum_{l=1}^N e^{\beta \epsilon_l}} 
\right)^{k_i} \\ \label{P2}
\left(     
1-\frac{e^{\beta \epsilon_i}}{\sum_{l=1}^N e^{\beta \epsilon_l}} 
\right)^{t-k_i}
\sum_{T(\{k\})} 
\prod_{l=1}^N \delta\left( g_l-\sum_{i=0}^{k_l}  A_{\rho,t}(T_{l,i}) \right).
\eea
Here $\{k\}=k_1, k_2, ...,k_n$ represents
a particular outcome of the game, in which site $1$ has been
visited $k_1$ times, site $2$ $k_2$ times, and so forth.
The symbol $T(\{k\})$ stands for the set of time sequences 
$T_{l,i}$ (the 
time step at which site $l$ has been visited for the
$i^{th}$ time) with a given set of $k$. There are 
$M(\{k\})=\frac{t!}{\prod_{i=1}^N k_i!}$
such sequences.

Having in hand the complete probability distribution \req{P2}
with coefficients \req{azero} and \req{aesse},
all quantities of interest can be calculated with projection techniques.

\subsection{Probability that the best wins}
As an example of quantities that can be calculated,
we shall find the probability $P_t^{1,1}$ that the site with 
the best
fitness has the greater number of visits at time $t$. Defining the events
$W_k=\left(g_k(t)>g_i(t) \; \forall i\neq k \right)$
and
$E_k=\left(\epsilon_k(t)>\epsilon_i(t) \; \forall i\neq k \right)$,
we can write
$$
P_t^{1,1}= \sum_{k=1}^N P_t(W_k | E_k) p_t(E_k) = N P_t(W_1,E_1).
$$
The joint probability above reads
\be \label{jp1}
P_t(W_1,E_1)= \int_0^1 dg_1 \prod_{i=2}^N \int_0^{g_1} dg_i 
\int_0^1 d\epsilon_1 \prod_{j=2}^N  \int_0^{\epsilon_1} d\epsilon_j P_t(\{ g_k \}).
\ee
If we now plug equation \req{P2}
into equation \req{jp1}, we obtain:
\bea \label{jp2}
P_t(W_1,E_1)&=& \sum^{1,N}_{k_1, k_2, ..., k_N; \sum_{i=1}^{N} k_i=t}
u_t(\{k\}) f_t(\{k\}) 
\\ \label{ut}
u_t(\{k\})&=& \sum_{T(\{k\})} \int_0^1 dg_1 \prod_{i=2}^N \int_0^{g_1} dg_i 
\prod_{l=1}^N \delta\left( g_l-\sum_{i=0}^{k_l}  A_{\rho,t}(T_{l,i}) \right)
\\ \label{ft}
f_t(\{k\})&=&  \int_0^1 d\epsilon_1 \prod_{j=2}^N  
\int_0^{\epsilon_1} d\epsilon_j
\prod_{i=1}^N 
\left( 
\frac{e^{\beta \epsilon_i}}{\sum_{l=1}^N e^{\beta \epsilon_l}} 
\right)^{k_i} 
\left(     
1-\frac{e^{\beta \epsilon_i}}{\sum_{l=1}^N e^{\beta \epsilon_l}} 
\right)^{t-k_i}.
\eea

While the integral in \req{ut} is straightforward, i.e.
\be \label{utsol}
u_t(\{k\})= \sum_{T(\{k\})} \prod_{l=2}^N \Theta\left[
\sum_{i=1}^{k_1} A_{\rho,t}(T_{1,i}) - 
\sum_{i=1}^{k_l} A_{\rho,t}(T_{l,i}) \right]
\ee
the one in equation \req{ft} needs some approximation to be 
solved. 

\subsubsection{Calculation of $f_t(\{k\})$}
Let us assume the $\epsilon_i$ are uniformly distributed between zero and one.
Defining the transformation of variables
\beas
y_i=e^{\beta \epsilon_i} / \sum_{l=1}^N e^{\beta \epsilon_l}\\
\Lambda=\sum_{l=1}^N e^{\beta \epsilon_l},
\eeas
equation \req{ft} can be written:
\be\label{ft1}
f_t(\{k\})= \frac{1}{N \beta} \int_N^{N e^{\beta}} d\Lambda \Lambda^{N-t-1}
\int_{1/N}^{\frac{e^{\beta}}{e^{\beta}+N-1 }} dy_1 
\int_{ \frac{1}{N e^{\beta}}  }^{y_1} \prod_{j=2}^N dy_j \prod_{i=1}^N
\left[ y_i^{k_i-1} (1-y_i)^{t-k_i} \right] 
\delta\left(1-\sum_{i=1}^N y_i\right).
\ee
Using the integral representation of the delta function
$$
\delta\left(1-\sum_{i=1}^N y_i\right)=
\int_{-i \infty}^{i \infty} dq e^{-q \left(1-\sum_{i=1}^N y_i\right)},
$$
one can evaluate the integral \req{ft1} by means of the saddle point method,
where the variable $q$ will play the role of a Lagrange multiplier.
Integrand maximization yields
\beas
{\tilde y}_i &=& \frac{k_i-1}{t-1} + \frac{q (k_i-1) (t-k_i)}{(t-1)^3}
\mbox{   for   } q \ll t-1\\
{\tilde q} &=& \frac{(N-1)(t-1)^2}{t^2-t N + t -\sum_{i=1}^N k_i^2};
\eeas
the former result is, therefore, valid only if  $t \gg N$. 
In this limit the approximate solution reads:
\be\label{sol1}
f_t(\{k\})=\frac{1}{t N^t \beta} \Theta(k_1-t/N) \prod_{i=1}^N 
{\tilde y}_i^{k_i-1} (1-{\tilde y}_i)^{t-k_i} \Theta(k_1-k_i).
\ee

In the opposite limit, that of large $N$, one can use the law of
large numbers over the fitness distribution, i.e.
$$
\sum_{l=1}^N e^{\beta \epsilon_l} \to
\Lambda = \frac{N}{\beta} (e^{\beta}-1) 
\; \textrm{for $\beta>0$}
$$
and
\bea \nonumber
\prod_{i=2}^N &&
\left( \frac{e^{\beta \epsilon_i}}{\sum_{l=1}^N e^{\beta \epsilon_l}} \right)^{k_i} 
\left( 1- \frac{e^{\beta \epsilon_i}}{\sum_{l=1}^N e^{\beta \epsilon_l}} \right)^{t-k_i} 
\to \\ \label{ct}
c_t (\{k\}) &=&
\exp\left[\sum_{i=2}^N \beta k_i <\epsilon_i> -t \log\Lambda +
(t-k_i) \avg{ \log \left( \Lambda-e^{\beta \epsilon_i} \right)} \right] \\
&=& \exp\left[\sum_{i=2}^N \frac{\beta k_i}{2} -t \log\Lambda +
\frac{(t-k_i)}{\beta} \sum_{l=1}^{\infty} \frac{e^{\beta l}-1}{\Lambda^l l^2} \right].
\eea
Here the angular brackets stand for averages over the distribution
of the $\epsilon_i$-s.
Hence:
\bea \label{ftmf}
f_t(\{k\}) &\simeq& c_t (\{k\}) \int_0^1 d\epsilon 
\left( \frac{e^{\beta \epsilon}}{\Lambda} \right)^{k_1}
\left( \frac{1-e^{\beta \epsilon}}{\Lambda} \right)^{t-k_1} 
\epsilon^{N-1} \\
&\simeq& c_t (\{k\})
\left( \frac{k_1}{t} \right)^{k_1}
\left( 1-\frac{k_1}{t} \right)^{t-k_1}
\left( \frac{\log \Lambda k_1/t}{\beta} \right)^{N-1}
\Theta\left( \Lambda k_1 -t \right),
\eea
where the saddle point method has been employed to solve the last
integral.

\section{Statistics of visits on a ultrametric argument space}
We now turn our attention to the entire World Wide Web. While inside a given
category it is easy to compare different sites, the WWW deals with various
arguments, sometimes not overlapping at all. Scalar fitnesses should, therefore,
be replaced by vectors
and players could only be active in the domains they are experts of.

Here we would like to introduce a simple hierarchical structure of
Internet categories, taking only into account passive players employing
different research efforts. Inspired by a recent work on social networks
\cite{Watts_social} we place $N=2^M$ sites on the leaves of a ultrametric tree
with $M$ levels. Upon labeling them sequentially from $0$ to $N-1$,
the ultrametric distance between two web pages $i$ and $j$ can be defined as
the greatest exponent $d$ such that
$$
[i/2^{d'}]=[j/2^{d'}]\mbox{   for   } d' \ge d,
$$
where $[a]$ denotes the integer part of $a$.
At each time step a visitor places a query (say $i \in [0,N-1]$) and extract distance 
$d$ from a distribution $\rho(d)$,
such that all the nodes within a radius $d$ from the query are eligible answers.
We assume that the probability that a site $j$ receives a visit as a consequence of
a given query $i$ is driven by the generalized preferential attachment rule
\cite{Krapivsky} 
$$
\Pi_j\propto\theta(d(i,j)-d)(n_j+1)^{\alpha}.
$$
Here the constant one represents the initial attractiveness of each node, which is
necessarily non-zero when there are no active players.
Since the system is not growing and the number of sites is fixed, 
as in the quasi-static scale free networks \cite{Manna}, simple 
linear preferential attachment is not enough for the number of visits to be power-law distributed.

\begin{figure}
\centerline{\epsfxsize=2.5in \epsfbox{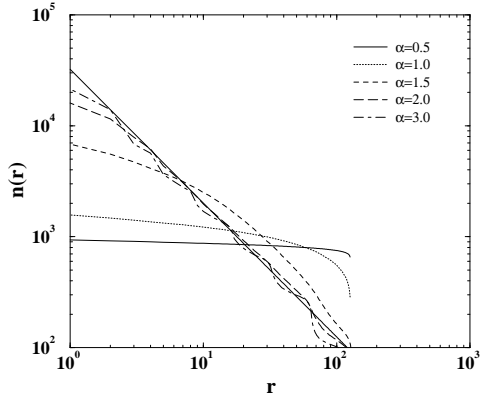}} 
\caption{Number of visits $n(i)$ versus rank $r_i$ in the ultrametric model
with uniform radius distribution,
for different values of $\alpha$. The simulations are performed on a system of $128$ webpages and the data are averaged over $100$ runs. For $\alpha = 2$ the best
fit to the data (indicated with solid line) is a power law
with exponent $\xi \simeq 1.2$. } 
\label{ultra_rank_pa.fig}
\end{figure}

Let us first analyze the model when $\rho(d)$ is uniform.
In figure $\ref{ultra_rank_pa.fig}$ 
the number of visits $n$ of a given web page $i$ is plotted
versus its rank $r(i)=N \sum_{n_i}^{\infty} P(n_i)$.
For $\alpha \geq 2$ we observe  a power-law behavior of $n(r)$
\be
n(r)\sim r^{-\xi} 
\label{xi}
\ee
corresponding to a power-law $P(n)\sim n^{-\frac{\xi+1}{\xi}}$ in the probability density of the number of visits.
For lower values of the exponent $\alpha$ (including the special case of linear preferential 
attachment) the power-law $(\ref{xi})$ disappears.

\subsection{Dependence from the specificity of the request}
We now analyze the role of distribution $\rho(d)$, which modulates the 
proportion of players who place a   very specific query with short distance $d$ and that
of players who are much more easily satisfied, looking for popular sites
within a wider radius.
In particular we  have considered the two distributions  
\bea\label{active.eq}
\rho(d)&=&(\kappa+1)(d/M)^{\kappa}\\ \label{passive.eq}
\rho(d)&=&(\theta+1)(1-d/M)^{\theta}.
\eea
In the first case queries that require a more precise answer (small distances $d$) 
are less frequent than queries that require a less precise answer. 
The reversal applies to the second case.

\begin{figure}
\centerline{\epsfxsize=3.5in \epsfbox{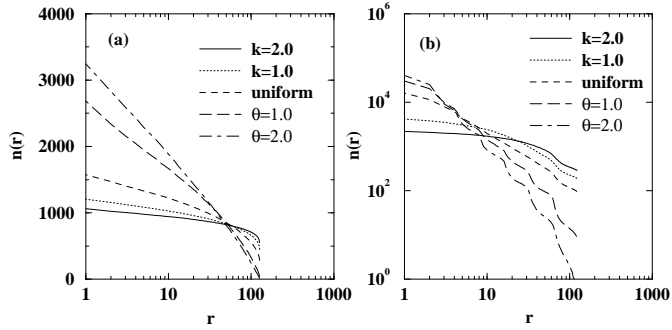}} 
\caption{Effect of the variation of $\rho(d)$ on the rank distribution of visits
in a system of $N=128$ nodes. In graph(a) the model studied has linear 
preferential attachment ($\alpha=1$), in graph(b) the model studied has $\alpha=2$.
Simulations are carried out for $T=10^5$ time steps. 
The two bottom lines are instances of distribution \req{active.eq},
the two top ones of \req{passive.eq}. The instance with uniform distribution
has also been drawn for comparison.} \label{a1.fig}
\end{figure}

We start with a system of web pages in which the visits dynamics is driven by a
preferential attachment with $\alpha=1$. The impact of the variation of  
distribution $\rho(d)$ is illustrated in Fig. $\ref{a1.fig}a$,
where the rank distribution of visits is displayed in linear scale.
Indeed the curve $n(r)$  becomes steeper as visitors increase their
search radius.

The role of distribution $\rho(d)$ becomes crucial
for greater values of $\alpha$.  
In particular in Fig. $\ref{a1.fig}b$ we consider the case $\alpha=2$.
For a distribution of the type $(\ref{active.eq})$ the power-law functional form of  
$n(r)$ (Eq. $\ref{xi}$) breaks down, leaving place to the logarithm of the rank $ n(r) \sim  \log(r)$.
For distribution \req{passive.eq} the scaling $(\ref{xi})$ is conserved, but the 
curves get steeper with increasing $\theta$ values.

\section{Conclusions}
We proposed a simple model of web pages attendance, focusing our attention  on webpages that are found in the same category. 
We investigated correlations between the number of visits they receive and their quality, as it emerges from the interplay of heterogeneous players: 
the passive players, driven by the popularity of the web page and its advertisement, and active players, driven by their own information
of the sites' intrinsic quality. We studied the model by numerical simulations and by analytical calculations.
Connectivity statistics follows power laws with different slopes, but
a typical length scale might occasionally appear, as in fig. 
\ref{cinquecento.fig}.
When a scalar indicator characterizes the intrinsic
quality of web sites, experts participation can improve the effectiveness
of Internet searches (fig. \ref{Pr_r} and \ref{P1,1t500.eps}).
In fact the  model displays a cross-over from the passive dominated phase to the active dominated phase,
in which the correlations between quality and visits build up.
It would be interesting to know where the actual Word-Wide-Web is placed
respect to this cross-over point and how much the popularity based ranking,
still used by some search engines,  
really reflects the ideal quality ranking of the web pages. 
 
We conclude the paper with a discussion of how the websites attendance varies in a system with multiple categories.
We formulate a hierarchical model in which we define a distance (the ultrametric distance) between different pages.
In this model there are no active players that visit their chosen web page but there are only passive players that 
visit web pages distant less than $d$ from a given query.
We have shown that the more generic is the search (the wider the radius $d$) the steeper is the distribution of 
the number of visits while the more specific is the search (the smaller is $d$) the smoother is the distribution of visits.

\section{Acknowledgments}
The authors would like to thank Matteo Marsili for helpful discussions.
This work was supported by the Swiss National Fund, Grant No. 20-61470.00.


\end{document}